\documentclass{emulateapj}

\slugcomment{Not to appear in Nonlearned J., 45.}

\begin{document}


\title{Self-similar structure of a hot magnetized flow with thermal conduction}


\author{M. Ghasemnezhad\altaffilmark{1} and M. Khajavi\altaffilmark{1}}
\affil{Department of Physics, School of Sciences, Ferdowsi University of Mashhad, Mashhad, 91775-1436, Iran}

\author{S. Abbassi\altaffilmark{2,3}}
\affil{School of Physics, Damghan University, PO Box 36715-364, Damghan, Iran}
\affil{School of Astronomy, Institute for Research in Fundamental Sciences (IPM), Tehran, 19395-5531, Iran}
\email{abbassi@ipm.ir}

\begin{abstract}
We have explored the structure of hot magnetized accretion flow with thermal conduction. The importance of thermal conduction in hot accretion flows has been confirmed by observations of the hot gas surrounding Sgr $A^*$ and a few other nearby galactic nuclei. For a steady state structure of such accretion flows a set of self similar solutions are presented. In this paper, we have actually tried to re-check the solution presented by Abbassi et al. (2008) using a physical constrain.  In this study we find that Eq 29 places a new constrain that limits answers presented by Abbassi et al. 2008. In that paper the parameter space in which it is established in the new constrain was plotted. However, the new requirement makes up only a small parameter space with physically acceptable solutions. And now in this manuscript we have followed the idea with more effort, and tried to find out how thermal conduction influences the structur of the disks in a physical parameter space. We have found out that the existence of thermal conduction will lead to reduction of accretion and radial and azimuthal velocities as well as the vertical thickness of the disk, which is slightly reduced. Moreover, the surface density of the disk will increase when the thermal conduction becomes important in the hot magnetized flow.
\end{abstract}



\section{Introduction}
It is widely believed that many astrophysical objects are powered by mass accretion on to black holes. Black hole accretion disk models have been extensively studied during the past three decades (see Kato et al. 2008 for a review). The standard geometrically thin, optically thick accretion disc model can be successfully explained by most of the observational features in active galactic nuclei (AGN) and X-ray binaries (Shakura \& Sunyaev 1973). Besides the traditional standard disk model by Shakura and Sunayev (1973), there are a number of new disk classifications including advection-dominated accretion flows (ADAF) or radiative inefficient accretion flows (RIAF) for a very small mass-accretion disks (Narayan \& Yi 1994) and supercritical accretion disk or so-called slim disk, for a very large mass-accretion rate (Abramowicz et al 1998).
ADAFs are optically thin and hot (compared to the virial temperature of the gas in the flows), which radiate mostly in X-ray waveband (see Narayan \& McClintock 2008). In the ADAF model, only a small fraction of the gravitational energy released in the accretion flow is radiated away due to inefficient cooling and most of it energy is stored in the accretion flow and advected to the black hole. This model can successfully explain the main observational features of black hole X-ray binaries and low-luminosity AGN (LLAGN) (e.g., Narayan \& Yi 1994, 1995a; Quataert et al. 1999; Yuan et al. 2003; Ho 2008).
The diversity of models indicates that modeling the hot accretion flows around a black hole is a challenging and controversial problem. One of the largely neglected physical phenomena in the modeling ADAFs is the thermal conduction. Recent observations of hot accretion flows around active galactic nuclei indicate that it should be based on collision-less regime (Tanaka \& Menou 2006). It is also suggested that accretion in these systems proceeds under the weakly collisional condition. Furthermore, it is proposed that thermal conduction can be a possible mechanism through which sufficient extra heating is provided in hot advection dominated accretion flows. So, thermal conduction seems to have an important role in energy transport in the accreting materials around a hot accretion disk where they are nearly completely ionized.
Since advection-dominated disks have high temperatures, the internal energy per particle is high. This is one of the reasons why advective cooling overcomes radiative cooling. For the same reason, turbulent heat transport by conduction in the radial direction is non-negligible in the heat balance equation. So it should be important to consider the role of thermal conduction in an ADAF solution. Shadmehri (2008), Faghei (2011), Abbassi et al. (2008, 2010) Tanaka \& Menou (2006) have studied the effect of hot accretion flow with thermal conduction in a semi-analytical method; physics of such systems have been studied in simulation models (e. g. Sharma et al. 2008; Wu et al. (2010). Shadmehri (2008) has shown that thermal conduction opposes the rotational velocity, but increases the temperature. Abbassi et al. (2008) have shown that for this problem there are two types of solutions; high and low accretion rates. They plotted the radial velocity for both solutions which revealed that it is influenced by thermal conduction. In this study, by introducing a new physical condition, it has been shown that high accretion rate solutions of Abbassi et al. (2008) are not exactly correct. Also some of the low accretion rate solutions presented there are not of physical value. With this extra condition it is possible to find the physically meaningful parametric space using which we next plotted the dynamical quantities to investigate how thermal conduction affects them.
Finally, tangled magnetic fields in accretion flows are likely to reduce the effective mean free path of particles. The magnitude of this reduction, which depends on field geometry, is still unknown. In this paper, we will investigate the effect of thermal conduction on the physical structure of ADAF-like accretion flow around a black hole in the presence of a toroidal magnetic field.
\section{The Basic Equations}
We are interested in analyzing the structure of a magnetized ADAF where thermal conduction plays an important role in energy transportation.  We only consider toroidal component of the magnetic field and suppose that the gaseous disk is rotating around a compact object of mass $ M_* $. Thus, for a steady axi-symmetric accretion flow, i.e. $\partial/ \partial t=\partial / \partial \phi=0$, we can write the standard equations in the cylindrical coordinates $ (r,\phi,z) $. In addition, we vertically integrate the flow equations and so, all the physical variables only become functions of radial distance $r$. Moreover, relativistic effects are neglected and Newtonian gravity in radial direction is taken into account. The disk is supposed to be turbulent and possesses an effective turbulent viscosity $\bf \nu$. We adopt $\alpha$-prescription for viscosity of a rotating gaseous disk. As for conservation of energy, it is assumed the energy generated due to viscosity dissipation is balanced by the radiation, thermal conduction and advection cooling (e.g., Narayan \& Yi 1994, Abbassi et al. 2008, 2010).
The equation of continuity reads
\begin{equation}
\frac{1}{r}\frac{\partial}{\partial r}(r\Sigma V_r )=2 \dot{\rho}H
\end{equation}
where $V_{r}$ is the accretion velocity ($V_{r}<0$) and
$\Sigma=2\rho H$ is the surface density at a cylindrical radius
$r$. Where $\dot{\rho}$ the mass loss rate per unit volume, H is the disk's half-thickness.
The equation of motion in the radial direction is:
\begin{equation}
v_r\frac{\partial v_r}{\partial r}=\frac{v_\varphi^{2}}{r}-\frac{G
M_{\ast}}{r^{2}}-\frac{1}{\Sigma}\frac{d}{dr}(\Sigma
c_s^{2})-\frac{c_A^{2}}{r}-\frac{1}{2\Sigma}\frac{d}{dr}(\Sigma
c_s^{2})
\end{equation}
where $v_{\varphi}, c_s$ and $c_A$ are the rotational velocity of the disk, sound velocity and Alfven velocities of the gas, respectively. Sound speed and the Alfven velocity are defined as $c_{\rm s}^2=p_{\rm gas}/ \rho$ and  $c_{\rm A}^2=B_{\varphi}^2 / 4\pi\rho=2p_{\rm mag}/ \rho$ where $p_{gas}$ and $p_{mag}$ are the gas and the magnetic pressures, respectively.

The vertically integrated angular momentum equation becomes:
\begin{equation}
r\Sigma V_r
\frac{d}{dr}(rV_\varphi)=\frac{d}{dr}(\frac{r^{3}\alpha c_{s}^{2}
\Sigma}{\Omega_{k}}\frac{d\Omega}{dr})
\end{equation}
where $\alpha$ is the viscose parameter,$\Omega(=\frac{v_{\varphi}}{r})$ the angular velocity and
$\Omega_{k}$ the Keplerian angular velocity, respectively.

By integrating over $z$ of the hydrostatic balance one gets:
\begin{equation}
\frac{GM}{r^{3}}H^{2}=c_s^{2}[1+\frac{1}{2}(\frac{c_A}{c_s})^{2}]=(1+\beta)c_s^{2}
\end{equation}
where $\beta=\frac{P_{mag}}{P_{gas}}=\frac{1}{2}(\frac{c_A}{c_s})^{2}$
which indicates the important of magnetic field pressure compared to gas pressure (Akizuki \& Fukue 2006, Abbassi et al. 2008, 2010). Two cases can be considered; Case 1: where the pressure is assumed to be the gas pressure (thermal pressure) (which is the choice of this paper) and Case 2: when the pressure is assumed to be the total of magnetic plus gas pressure. So in case 2 we replace $\alpha$ with $\alpha(1+\beta)$ in all of the equations.
Next, the dynamical properties of the disk for different values of $\beta$ is demonstrated. The energy equation considering cooling and heating processes in an ADAF can be found as follows. We assume the energy generated due to viscous dissipation and the heat conducted into the volume are balanced by the advection cooling and energy transport by thermal conduction. Thus,
\begin{equation}
\frac{\Sigma V_r}{\gamma-1}\frac{dc_s^{2}}{dr}-2H V_r
c_s^{2}\frac{d\rho}{dr}=\frac{f\alpha\Sigma
c_s^{2}}{\Omega_k}r^{2}(\frac{d\Omega}{dr})^{2}-\frac{2H}{r}\frac{d}{dr}(r^{2}F_s)
\end{equation}
On the right-hand side of the energy equation we have:
\begin{displaymath}
Q_{+}-Q_{-}+Q_{cond}=Q_{f}+Q_{cond}
\end{displaymath}

where $Q_{cond}=-\nabla\cdot F_s$ represents energy transfer due to the thermal conduction and $F_s=5\Phi_s \rho c_s^{3}$ is the saturated conduction flux ( Cowie \& Mackee1977) in the direction of temperature gradient. $\Phi_s$, the dimensionless coefficient, is less than unity.
Finally since we consider the toroidal component for the global magnetic field, the induction equation can be written as:
\begin{equation}
\frac{d}{dr}(V_r B_\varphi)=\dot{B_\varphi}
\end{equation}

where $\dot{B_\varphi}$ is the field scaping/creating rate due to magnetic instability or dynamo effect.
Now we have a set of MHD equations that describe the structure of magnetized ADAFs. The solutions to these equations are strongly correlated to viscosity, magnetic field strength $\beta$, the degree of advection $f$ and the thermal conduction parameter $\phi_s$ for the disks. We seek a self-similar solution for the above equations. In the next section we will present self-similar solutions to these equations.

\section{Self-Similar Solutions}
In order to have a better understanding of the physical processes taking place in our disks, we seek self-similar solutions of the above equations. The self-similar method has a wide range of applications for the full set of MHD equations although it is unable to describe the global behavior of accretion flows since no boundary conditions have been taken into account. However, as long as we are not interested in the behavior of the flow near the boundaries, these solutions are still valid.
In the self-similar model the velocities are assumed to be expressed as follows:
\begin{equation}
V_r(r)=-c_1 \alpha V_k(r)
\end{equation}
\begin{equation}
V_\varphi(r)=c_2 V_k(r)
\end{equation}
\begin{equation}
c_s^{2}=c_3 V_k^{2}
\end{equation}
\begin{equation}
c_A^{2}\frac{B_\varphi^{2}}{4\pi\rho}=2\beta c_3 \frac{GM}{r}
\end{equation}
where
\begin{equation}
V_k(r)=\sqrt{\frac{GM}{r}}
\end{equation}
and constant $c_1$,$c_2$ and $c_3$ are determined later from the main MHD equations. We obtain the disk half-thickness H as:
\begin{equation}
\frac{H}{r}=\sqrt{c_3 (1+\beta)}=\tan\sigma
\end{equation}
Hence, a hot accretion also has a conical surface, whose opening (half-thickness) angle is $\sigma$.
Assuming the surface density $\Sigma$ to be in the form of:
\begin{equation}
\Sigma=\Sigma_0 r^{s}
\end{equation}
we obtained,e.g.
\begin{equation}
\dot{\rho}=\dot{\rho_0} r^{s-\frac{5}{2}}
\end{equation}
\begin{equation}
\dot{B_\varphi}=\dot{B_0} r^{\frac{s-5}{2}}
\end{equation}

By substituting the above self-similar solutions in to the the equations of the disks, we obtain the following system of dimensionless equations, to be solved for $c_1$,$c_2$ and $c_3$: In case 1:
\begin{equation}
\dot{\rho_0}=-(s+\frac{1}{2})\frac{c_1\alpha\Sigma_0)}{2}\sqrt{\frac{GM_\ast}{(1+\beta)c_3}}
\end{equation}
\begin{equation}
H=\sqrt{(1+\beta)c_3}r
\end{equation}
\begin{equation}
-\frac{1}{2}c_1^{2}\alpha^{2}=c_2^{2}-1-[s-1+\beta(s+1)]c_3
\end{equation}
\begin{equation}
c_1=3(s+1)c_3
\end{equation}
\begin{equation}
(\frac{1}{\gamma-1}-\frac{1}{2})c_1 c_3 =\frac{9}{4}f c_3
c_2^{2}-\frac{5 \Phi_s}{\alpha}(s-\frac{3}{2})c_3^{\frac{3}{2}}
\end{equation}

It is evident from the above equations that for $s = -1/2$ there is no mass loss while it is present for the case where $s > -1/2$. On the other hands, the escape and creation of magnetic fields balance one another for $s = 3$. Although outflow is one of the most important processes in accretion theory,(see Narayan \& Yi 1995; Blandford \& Begelman 1999; Stone \& Pringle \& Begelman1999 and also some recent work like Xie \& Yua 2008; Ohsuga \& Mineshige 2011), but in our model we choose a self-similar solution in which $\dot{\rho}=0$ and $\dot{B}_{\phi}\propto r^{-11/4}$ ($s = -1/2$) thus ignoring the effect of wind and outflow on the structure of the disk.
After some algebraic manipulations, we obtain a forth order algebraic equation for $c_1$:
\begin{equation}
D^{2}c_1^{4}+2DBc_1^{3}+(B^{2}-2D)c_1^{2}-(2B+A^{2})c_1+1=0
\end{equation}

where the coefficients depend on the input parameter as:
\begin{equation}
D=\frac{1}{2}\alpha^{2}
\end{equation}
\begin{equation}
B={\frac{4}{9f}(\frac{1}{\gamma-1}-\frac{1}{2})-[s-1+\beta(s+1)][\frac{1}{3(s+1)}]}
\end{equation}
\begin{equation}
A=\frac{20\Phi_s}{9f\alpha}(s-\frac{3}{2})[\frac{1}{3(s+1)}]^{\frac{1}{2}}
\end{equation}

This algebraic equation shows that the variable $c_1$ which determines the behavior of radial velocity only depends on the $\alpha$, $\Phi_s$, $\beta$ and $f$. Using $c_1$ from this equation, the other variables (i.e. $c_2$ and $c_3$) can be determined easily:
\begin{equation}
c_2^{2}=\frac{4 c_1}{9f}[\frac{1}{\gamma-1}-\frac{1}{2}]
+\frac{20\Phi_s}{9f\alpha}(s-\frac{3}{2})[\frac{1}{3(s+1)}]^{\frac{1}{2}}
c_1^{\frac{1}{2}}
\end{equation}
\begin{equation}
c_3=c_1(\frac{1}{3(s+1)})
\end{equation}

As was stated above we are solving the problem in the case where there is no wind, $s=-1/2$. When $s=-\frac{1}{2}$ the expressions for $c_1$ and $c_2$ are found to be:
\begin{equation}
c_2^{2}=\frac{2c_1}{9f}(\frac{3-\gamma}{\gamma-1})-\frac{40}{9\alpha f}\sqrt{\frac{2}{3}}\phi_s\sqrt{c_1}
\end{equation}
\begin{equation}
c_3=\frac{2}{3}c_1
\end{equation}

Abbassi et al. (2008) have found two types of solutions which represent high and low accretion rates. These two rates act differently towards the saturated thermal conduction parameter, $\phi_s$. As it was stated in the introduction the main aim of this paper is to place an additional physical condition. This new requirement limits the solutions of $c_2^{2}\geq 0$. From the above equation, $c_2^{2}\geq 0$ and $\gamma=\frac{5}{3}$, we have:

\begin{equation}
\sqrt{c_1}\geq \frac{1}{\alpha}\sqrt{\frac{200}{3}}\phi_s
\end{equation}
Therefore the $(\phi_s,\alpha)$ parametric space could be found as follows:

\begin{equation}
\alpha-\sqrt{\frac{200}{3c_1}}\phi_s\geq 0
\end{equation}
A numerical solution to this equation is presented in Fig. 1 for variety of values of the viscosity parameter $\alpha$ and the thermal conduction parameter $\phi_s$. These solutions may be considered either as flows with a fixed value of viscosity parameter and a sequence of increasing thermal conduction parameter or a sequence of different values of viscosity parameter $\alpha$ in a fixed thermal conduction regime.
These plots show that the new physical condition places a constraint on the physically valid parametric space. As a result we used this extra condition to check the validity of the solutions presented by Abbassi et al. (2008).
Tanaka \& Meneu (2006), Abbassi et al. (2008) have shown that for a very small $\phi_s$ their solutions coincide the original 1D ADAF solutions; but by adding the saturated conduction parameter, $\phi_s$, the effect of thermal conduction can be better seen while approaching $~0.001-0.01$. On the other hand, the widely accepted values of $\alpha$ are between $0.01-0.1$. So we have to plot our solutions in this range. The dark areas in Fig1 are where we are looking for; regions of the parametric space physically acceptable.
Another aim of this study was tore-check the solution presented by Abbassi et al. 2008 and and further to study the effect of heat conduction and magnetic field on the structure of ADAFs.
Fig. 1 shows that for larger values $\alpha$ we should choose a large $\phi_s$ in order to be sure to have a physically valid solution. The calculations are carried out in a range of viscosities, thermal conduction and magnetic field parameters. The calculated global structures of some ADAFs with different parameters are plotted in Figures 2-8.
\input{epsf}
\epsfxsize=3.1in \epsfysize=2.5in
\begin{figure}
\centerline{\epsffile{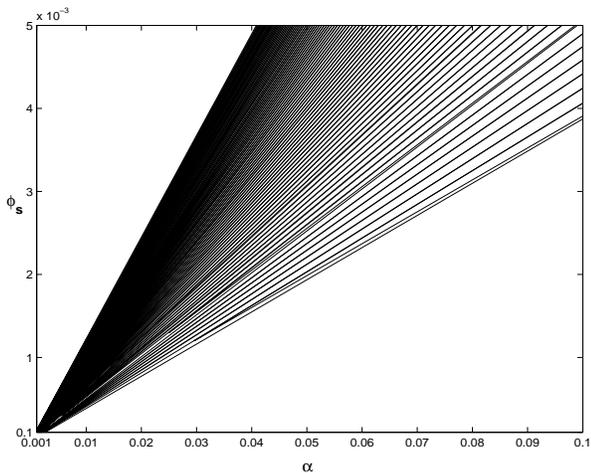}}
 \caption{The parameter space for thermal conduction parameter $\phi_s$ as a function of viscous parameter $\alpha$.
The dark section of the plot shows the physically accepted area of the parametric space}.
 \label{figure1 }
\end{figure}
\input{epsf}
\epsfxsize=3.1in \epsfysize=3.in
\begin{figure}
\centerline{\epsffile{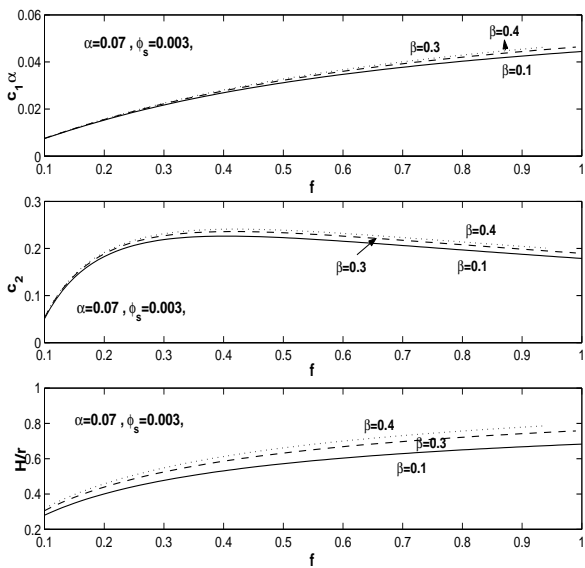}}
 \caption{Numerical coefficient $C_i$s as a function of advection parameter $f$ for several values of $\alpha$ viscose parameter.All of this figures was set up for $\alpha=0.07$,$\phi_s=0.003$}.
\label{figure2}
\end{figure}
\input{epsf}
\epsfxsize=3.1in \epsfysize=3.in
\begin{figure}
\centerline{\epsffile{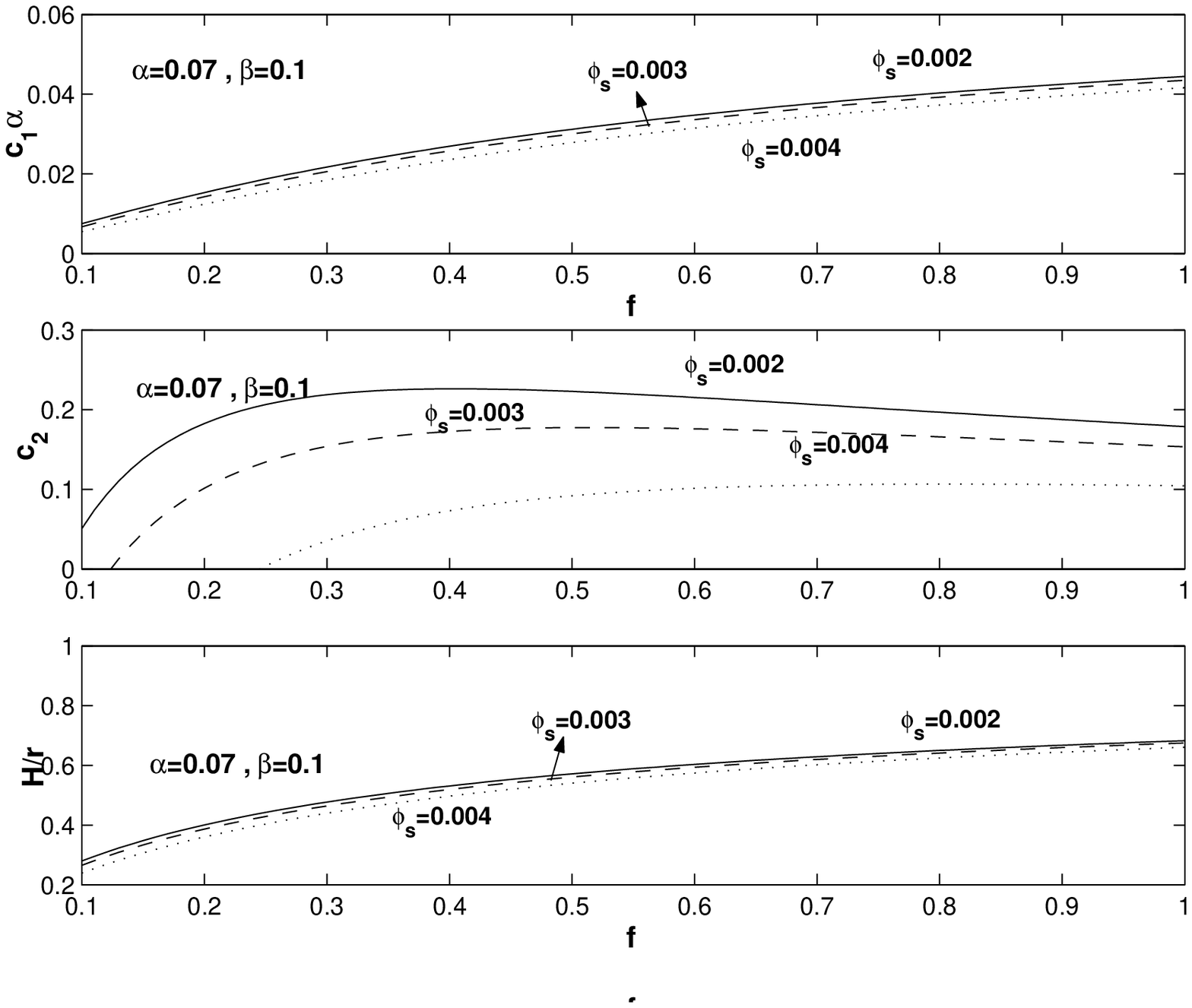}}
 \caption{Numerical coefficient $C_i$s as a function of advection parameter $f$ for several values of $\phi_s$ thermal conduction parameter.All of this figures was set up for $\alpha=0.07$,$\beta=0.1$}.
\label{figure3}
\end{figure}\input{epsf}
\epsfxsize=3.1in \epsfysize=3.in
\begin{figure}
\centerline{\epsffile{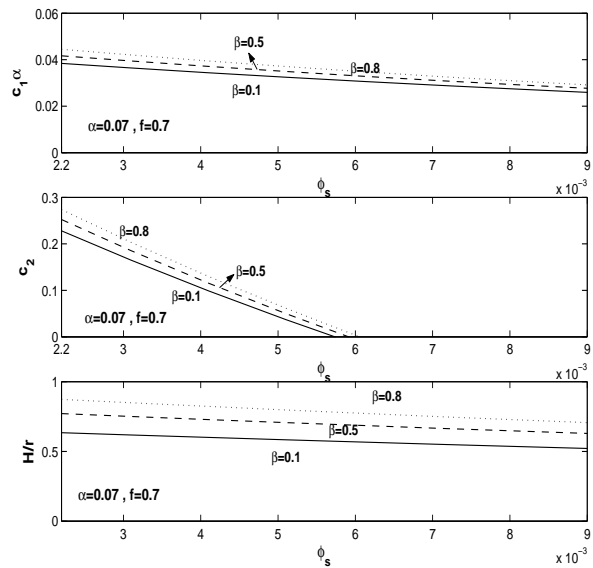}}
\caption{Numerical coefficient $C_i$s as a function of thermal conduction parameter $\phi_s$ for several values of $\beta$ magnetic field parameter All of this figures was set up for $\alpha=0.07$,$f=0.7$}.
 \label{figure4}
\end{figure}
\input{epsf}
\epsfxsize=3.1in \epsfysize=3.in
\begin{figure}
\centerline{\epsffile{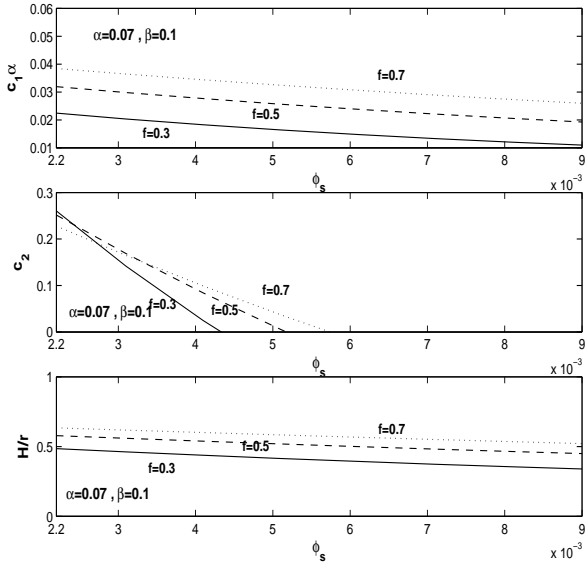}}
\caption{Numerical coefficient $C_i$s as a function of thermal conduction parameter $\phi_s$ for several values of $f$ advection parameter.All of this figures was set up for $\alpha=0.07$,$\beta=0.1$}.
\label{figure5}
\end{figure}

Figure 2 shows how the coefficient $C_is$ depend on the magnetic field parameter, $\beta$ for different values of advection parameter $f$. Radial velocity is determined by C1 which is shown in the upper panel. In ADAFs the radial velocity is generally less than free fall velocity on a point mass, but it becomes larger if the advection parameter $f$ is increased and such behavior is consistent with the previous analytical solutions (e.g., Akizuki \& Fukue 2006; Abbassi et al. 2008, 2010).
For a given $f$, both the radial and rotational velocities increase as the disk becomes more magnetized (i.e., larger $\beta$). A magnetized disk must rotate faster than the case where no magnetic field is present because of the effect of magnetic tension. It is clear that increasing the magnetic field strength increases radial velocity (accretion rate) in the accretion flows. Radial velocity of the disk is more sensitive to the variations of $f$ when the disk is more magnetized. The rotational velocity behaves the same for different values of $f$ but it slightly shifts up when parameter $\beta$ increases. The standard ADAFs becomes thicker when the disks is advective $f\sim1$ but by adding the magnetic field strength the vertical thickness is increased significantly for a given $f$.
In Fig. 3 we investigate the role of saturated thermal conduction in radial, toroidal and the vertical thickness of accretion flows. Increasing the thermal conduction coefficient $\phi_s$ will decrease the radial velocity. It will also have a large effect on the rotational velocity of accretion flows (middle panel). Furthermore it will decrease the sound speed and therefore vertical thickness of the disks too (lower panel).
In Figs 4 and 5 we have plotted the coefficient Ci with thermal conduction parameter $\phi_s$ for different values of $\beta$ and f. Tanaka \&Menou (2006) have shown that for a very small $\phi_s$ their solutions coincide the original ADAF solution (Narayan \& Yi 1994), but by adding the saturated conduction parameter, $\phi_s$, the effect of thermal conduction can be better seen while approaching $~0.001-0.01$. On the other hand, the widely accepted values of $\alpha$ are between $0.01-0.1$. So we have to plot our solutions in this range. We can see that for a given set of input parameters, the solution reaches a non-rotating limit at a specific value of $\phi_s$. We cannot extend the solution for values of this limit for $\phi_s$ because equation (25) gives a negative value for $C_2^2$, which is clearly not valid. The values of this limitation on $\phi_s$ increases by adding magnetic field parameter, $\beta$, fig 4 (middle panel), advection parameter, $f$, fig 5 (middle panel) smoothly.  In fact the higher values of $\beta, f$ will allow the disk to have a physical rotating solution for larger $\phi_s$. As it is seen in the lower panels of figures 4 and 5 by increasing $\beta, f$ the vertical thickness of the disks in a given $\phi_s$ will increase. But the vertical thickness in a fixed $\beta$ or $f$, will decrease globally as $\phi_s$ increased. It would be expected that in high $\phi_s$, disks becomes thin as it was predicted by Abbassi et al (2010) and Shadmehri (2008).
\input{epsf}
\epsfxsize=3.1in \epsfysize=3.in
\begin{figure}
\centerline{\epsffile{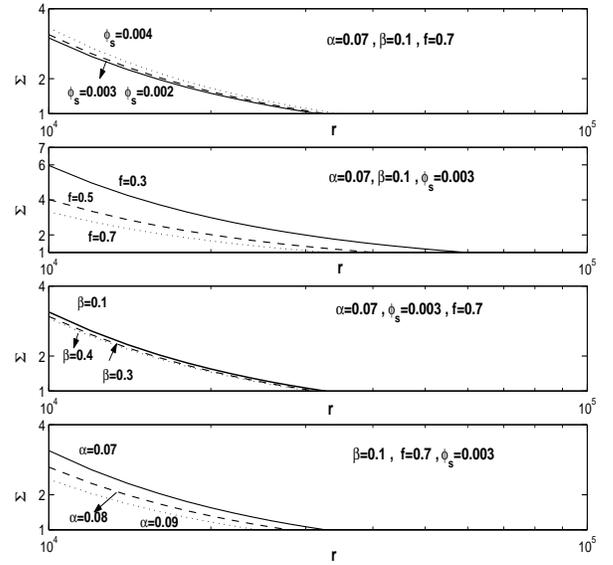}}
\caption{Surface density $\Sigma (kg/m^{2})$ as a function of the radius ($r$)of disk for several values of $\phi_s , f , \beta ,
 \alpha$ parameters}.
\label{figure6}
\end{figure}
\input{epsf}
\epsfxsize=3.1in \epsfysize=3.in
\begin{figure}
\centerline{\epsffile{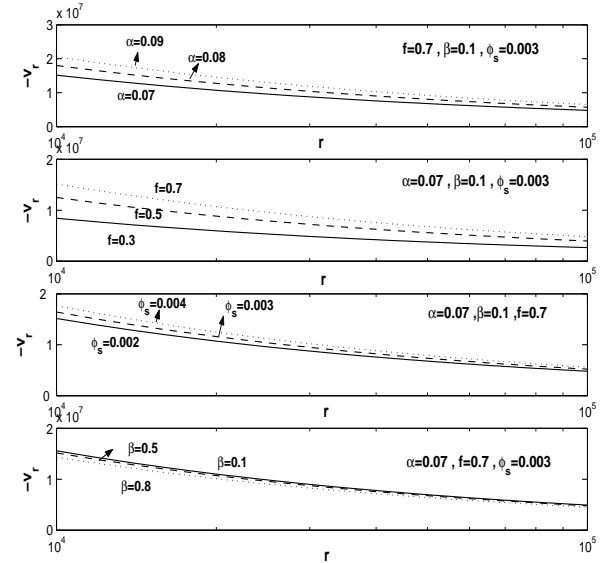}}
\caption{Radial velocity $v_r (m/s)$ as a function of the radius ($r$)of disk(in unit of $r_g$)for several values of $\phi_s , f , \beta ,
\alpha$ parameters}.
\label{figure7}
\end{figure}
As the level of thermal conduction is increased, more heat flows outward from the hot inner regions, resulting in a local increase of the gas temperature relative to the original ADAF solutions. Simultaneously, the gas adjusts its angular velocity and increases its inflow speed to conserve its momentum balance. We have found that the level of advection, $f$, will significantly modify the structure of the disk. The breakdown of the solutions, when $C_2^2\rightarrow0$, occurs at lower values of $\phi_s$ . Clearly thermal conduction can significantly affect the structure and the properties of the hot accretion flows. In order to understand the role of thermal conduction better it is useful to study it in a more realistic two-dimensional case, without height integrations. But at this stage we are going to find out how thermal conduction affects the structure of the disks with a simple analysis. It would be interesting to show how the surface density, radial velocity and accretion rate which are all observables, change with the radius of the disk when the thermal conduction plays an important role.
To show how surface density is affected by our input parameters we use:
\begin{equation}
\dot{M}=2\pi r \Sigma (-v_{r})
\end{equation}
where accretion rate is constant from the continuity equation; Eq(1). Assuming $\dot{M}\approx 10^{-8}M_{\odot}yr^{-1}$ and $M=10 M_{\odot}$ and by using self-similar solution for $v_{r}$ we have:
\begin{equation}
\Sigma=\frac{\dot{M}}{2 \pi C_1 \alpha \sqrt{GM}}r^{-\frac{1}{2}}
\end{equation}
As it is seen in figures 6 and 7, we have plotted the surface density, $\Sigma$, and radial velocity as a function of $r$. As expected, the absolute values of surface density and radial velocity will decrease as one moves out farther in the disk. In figure 6 it is evident that surface density is increased when thermal conduction plays an important role. But by adding the advection parameter $\Sigma$ will decrease significantly.
By adding viscous and advection parameters the absolute values of radial velocity will increased in a fixed $r$, as expected. Thermal conduction has the same effect and causes high radial velocity when it has an important role. Compared to other physical parameters, the magnetic field, $\beta$, does not have a significant effect on surface density and velocity gradients.

\section{SUMMERY AND CONCLUSION}
In this paper, we studied an accretion disk in the advection dominated regime by considering a purely toroidal magnetic field and in presence of thermal conduction.
Some approximations were made in order to simplify the main equations. We assumed an axially symmetric, static disk with the $\alpha$-prescription of viscosity. We also ignored relativistic effects and self-gravity of the disk. Considering the weakly collisional nature of a hot accretion flow (Tanaka \& Menou 2006; Abbassi et al. 2008), a saturated form of thermal conduction was adopted as a possible mechanism for energy transportation. We have accounted for this possibility by allowing the saturated thermal conduction constant, $\phi_s$, to vary in our solutions.
A set of similarity solutions was presented for such a configuration. Also following those presented by Abbassi et al. (2008) we re-examined the validity of their solutions by putting an additional physical condition. They had found two types of solutions: low and high accretion rate solutions. Their solutions had different behaviors towards thermal conduction parameters. We found out that the high accretion rate solution presented by Abbassi et al. (2008) is not physically valid. Also our new imposed condition puts some physical constrain on low accretion rate solutions.
As a result in order to have a physically valid solution only a small part of parameter space should be considered. After we assuring the validity of our solutions, by using proper values of input parameter, $\alpha, \beta, \phi_s, f$, we investigated the influence of thermal conduction on the structure of accretion flows. It is shown that the surface density is increased when the thermal conduction plays an important role. Also thermal conduction has the same effect on radial velocity.
However, there exist some limitations to our solutions. One is that the self-similar hot accretion flow with conduction is a single-temperature structure. Thus if one uses a two-temperatures structure for the ions and electrons in the disks, it is expected that the ions and electron temperatures decouple in the inner regions, which will modify the role of conduction. The other limitation of our solution is the anisotropic character of conduction in the presence of magnetic field. Balbus (2001) has argued that the structure of the hot flows could be affected by the anisotropic character of thermal conduction in the presence of magnetic field.
Although our preliminary self-similar solutions are too simplified, they clearly improve our understanding of the physics of ADAFs around a black hole. To have a realistic picture of an accretion flow a global solution is needed rather than the self-similar one. In our future studies we intend to investigate the effect of thermal conduction on the observational appearance and properties of a hot magnetized flow.
\acknowledgments

\clearpage

\end{document}